\documentstyle[12pt]{article}
\renewcommand{\baselinestretch}{1.75}

\newcommand{\ep}{\epsilon}

\newcommand{\th}{\theta}

\newcommand{\la}{\lambda}

\newcommand{\La}{\Lambda}

\newcommand{\be}{\begin{eqnarray}}
\newcommand{\ee}{\end{eqnarray}}

\newcommand{\lra}{\longrightarrow}

\newcommand{\llra}{\longleftrightarrow}
\newcommand{\pr}{\partial}

\newcommand{\np}{\newpage}
\newcommand{\hs}{\hspace}
\newcommand{\vs}{\vspace}

\newcommand{\nn}{\nonumber}

\newcommand{\bLa}{L}

\begin{document}

\thispagestyle{empty}

\vs*{-25mm}
\begin{flushright}
BRX-TH-436\\[-.2in]
BOW-PH-112\\[-.2in]
HUTP-98/A045 \\[-.2in]
hep-th/9806144
\end{flushright}

\begin{center}
{\Large{\bf One-instanton predictions  
for non-hyperelliptic curves 
derived from M-theory}} \\
\vspace{.2in}

\renewcommand{\baselinestretch}{1}
\small
\normalsize
Isabel P. Ennes\footnote{Research supported 
by the DOE under grant DE--FG02--92ER40706.}\\
Martin Fisher School of Physics\\
Brandeis University, Waltham, MA 02254

\vspace{.1in}
Stephen G. Naculich\footnote{Research supported 
in part by the National Science Foundation under grant no. PHY94-07194
through the ITP scholars program.}\\
Department of Physics\\
Bowdoin College, Brunswick, ME 04011

\vspace{.1in}

Henric Rhedin\footnote{Supported by the Swedish Natural
Science Research Council (NFR),  grant no.
F--PD1--883--305.}\\
Martin Fisher School of Physics\\
Brandeis University, Waltham, MA 02254

\vspace{.1in}

Howard J. Schnitzer\footnote{Research supported in part
by the DOE under grant DE--FG02--92ER40706.}\\
Lyman Laboratory of Physics\\
Harvard University, Cambridge, MA 02138\\
and\\
Martin Fisher School of Physics\footnote{Permanent address.\\
{\tt \phantom{aaa} naculich@bowdoin.edu; 
ennes,rhedin,schnitzer@binah.cc.brandeis.edu}}\\
Brandeis University, Waltham, MA 02254

\vspace{.2in}

{\bf{Abstract}} \end{center}
\renewcommand{\baselinestretch}{1.75}
\small
\normalsize
\begin{quotation}
\baselineskip14pt
\noindent One-instanton predictions are obtained from certain 
non-hyperelliptic Seiberg-Witten curves derived from M-theory 
for N=2 supersymmetric gauge theories. 
We consider SU$(N_1)\times$SU$(N_2)$ gauge theory
with a hypermultiplet in the bifundamental representation 
together with hypermultiplets in the defining representations 
of SU$(N_1)$ and SU$(N_2)$.
We also consider SU$(N)$ gauge theory with 
a hypermultiplet in the symmetric or antisymmetric representation, 
together with hypermultiplets in the defining representation. 
The systematic perturbation expansion about a hyperelliptic curve 
together with the judicious use 
of an involution map for the curve of the product groups 
provide the principal tools of the calculations. 
\end{quotation}

\np 

\setcounter{page}{1}

\noindent{\bf 1. Introduction}

The Seiberg-Witten approach \cite{SeibergWitten1} 
to deriving the exact low-energy properties of N=2 
supersymmetric gauge theories depends on the following data: a curve, 
which for many cases represents a Riemann surface, and a preferred 
meromorphic one-form, the Seiberg-Witten (SW) differential $\la$. When 
dealing with a Riemann surface, one calculates the renormalized 
order parameters of the theory, and their duals, from 
\be
2\pi i a_k=\oint_{A_k}\la \hs{10mm} {\rm and} \hs{10mm}
2\pi i a_{D,k}=\oint_{B_k}\la,\label{orderpar}
\ee
respectively, where $A_k$ and $B_k$ are a canonical basis of 
homology cycles for the Riemann surface. Given (\ref{orderpar}), 
the prepotential ${\cal F}$ is obtained by integrating 
\be
a_{D,k}=\frac{\pr{\cal F}}{\pr a_k}. \label{adfrel}
\ee
In terms of N=1 superfields, the Wilson effective Lagrangian, 
to lowest order in the momentum expansion, is 
\be
{\cal L}=\frac{1}{4\pi}{\rm Im}\left[\int{\rm d}^4\th
\frac{\pr{\cal F}(A)}{\pr A^i}\bar{A}^i+
\frac{1}{2}\int{\rm d}^2\th \frac{\pr^2{\cal F}(A)}{\pr A^i\pr A^j}
W^iW^j\right],
\ee
where $A^i$ are N=1 chiral superfields. Holomorphy implies that 
the prepotential of the Coulomb phase has the form 
\be
{\cal F}(A)={\cal F}_{cl.}(A)+{\cal F}_{1-loop}(A)+
\sum_{d=1}^{\infty}\La^{[I(G)-I(R)]d}{\cal F}_{d-inst.}(A),
\label{prepot}
\ee
where $I(G)$ is the Dynkin index of the adjoint representation 
of the gauge group, $I(R)$ is the sum of the Dynkin indices 
of the matter hypermultiplets, with the sum in (\ref{prepot}) being 
over the instanton expansion, and $\La$ is the quantum scale. 

A wide class of SW problems can be solved by means of hyperelliptic 
curves \cite{Everybody}, for which methods to extract the instanton 
expansion, 
as well as strong coupling information, have been rather well developed 
\cite{DHokerKricheverPhong1,DHokerKricheverPhong2,IsidroEdelsteinMarinoMas}. 
However, not all SW problems lead to hyperelliptic 
curves. 
In particular, 
M-theory \cite{Mtheory}--\cite{LL} and geometric engineering 
\cite{geomenginering} often lead to Riemann surfaces 
that are not hyperelliptic, and to varieties that are not Riemann surfaces 
at all. Methods required to extract {\it explicit} predictions 
for the prepotentials associated  to non-hyperelliptic curves are just 
beginning to be developed. In particular, in two 
previous papers \cite{NaculichRhedinSchnitzer,EnnesNaculichRhedinSchnitzer}, 
we provided a construction for the instanton expansion for SU($N$) 
gauge theory, with matter in the antisymmetric or symmetric 
representations, with explicit results for the one-instanton 
contribution to the prepotential. (These two cases involve 
non-hyperelliptic cubic curves.) It is extremely important to 
continue the program of computing {\it explicit} predictions for 
the prepotentials from M-theory and geometric engineering, and then 
checking these against independent microscopic field theoretic 
calculations \cite{instanton,KohzeMattisSlater}. 
Such successful comparisons should increase our 
confidence in the ability of string theory to provide us with field 
theoretic information, and in the power of M-theory and geometric 
engineering. 

This paper extends our work to three more 
cases, all involving non-hyperelliptic curves. We will consider 
the instanton expansions for 
(1) SU$(N_1)\times$SU$(N_2)$ gauge theory 
with one multiplet in the bifundamental representation, together 
with $K_0$ and $K_3$ hypermultiplets in the defining representation 
of SU$(N_1)$ and SU$(N_2)$ respectively, 
(2) SU$(N)$ gauge theory with 
one matter hypermultiplet in the symmetric 
representation, together with $N_f$ hypermultiplets in the defining 
representation, 
and (3) SU$(N)$ gauge theory with 
one matter hypermultiplet in the antisymmetric 
representation, together with $N_f$ hypermultiplets in the defining 
representation.

\noindent{\bf 2. SU$(N_1)\times$SU$(N_2)$ gauge theory}

Consider the N=2 supersymmetric gauge theory based on the
gauge group SU$(N_1)\times$ SU$(N_2)$,
with one massless hypermultiplet 
in the $(N_1,\bar{N_2})$ bifundamental representation, 
together with $K_0$ and $K_3$ massless hypermultiplets 
in the defining representation of SU$(N_1)$ and SU$(N_2)$ respectively. 
The chiral multiplets in the adjoint representation of SU($N_1$) 
or SU($N_2$) contain a complex scalar field $\phi_1$ or $\phi_2$. 
Along the flat directions of the potential,  
$[\phi_j,\bar{\phi}_j]$ vanishes $(j=1,2)$, 
and the symmetry is broken to U$(1)^{N_1-1}\times $U$(1)^{N_2-1}$.
The $(N_1-1)+(N_2-1)$ dimensional 
moduli space is parametrized classically by 
$e_i$ ($1\leq i\leq N_1$) and 
$\hat{e}_i$ ($1\leq i\leq N_2$), 
which are the eigenvalues of $\phi_1$ and $\phi_2$ respectively, 
and satisfy the constraints
$\sum_{i=1}^{N_1} e_i = 0$ and   
$\sum_{i=1}^{N_2} \hat{e}_i = 0$.

The curve for this theory, derived by Witten 
\cite{Witten1}, and made more explicit  in ref. \cite{ErlichNaqviRandall} 
is 
\be
P_0(x)\,t^3\,-\,{P_1(x)\over L_1^2}\,t^2\,+\,{P_2(x)\over L_1^2}\,t\,-\,
{L_2^2\,P_3(x)\over L_1^2}\,=0, \label{one}
\ee
where 
\be
P_0(x)\,&=&\,x^{K_0},\,\,\,\,\,\,\,\,\,\,
P_1(x)\,=\prod_{i=1}^{N_1}\,(x-e_i)\cr
P_2(x)\,&=&\prod_{i=1}^{N_2}\,(x-\hat{e}_i), \,\,\,\,\,\,\,\,\,
P_3(x)\,=\,x^{K_3},\cr
\vphantom{\left[{L_1^2\over 8}\right]}
L_1^2\,&=&\,\Lambda^{2N_1-N_2-K_0}_1\,,\,\,\,\,\,\,\,\,\,
L_2^2\,=\,\Lambda^{2N_2-N_1-K_3}_2, \label{two}
\ee
with $\La_1$ and $\La_2$ the quantum scales of the two gauge groups. 
The requirement of asymptotic freedom, and restriction to the Coulomb 
phase, implies that $\La_1$ and $\La_2$ appear with positive powers 
in (\ref{two}). 
The change of variables $t\,=\,y/(P_0(x)\,L_1^2)$ gives the curve 
\be
y^3\,-\,P_1(x)\,y^2\,+\,L_1^2\,P_0(x)\,P_2(x)\,y\,-
L_1^4\,L_2^2\,P_0^2(x)\,P_3(x)\,=\,0, \label{four}
\ee
in the form which we will analyze. 
The involution map 
\be 
y\rightarrow {L_1^2\,L_2^2\,P_0\,P_3\over y}
 \label{five}
\ee
interchanges the order of the gauge groups, i.e. 
SU$(N_1)\times $SU$(N_2)\lra $SU$(N_2)\times $SU$(N_1)$;
this will be important in what follows. 
The curve (\ref{four}) corresponds to a three-fold 
branched covering of the 
Riemann sphere, with sheets one and two connected by $N_1$ 
square-root branch-cuts centered about $x=e_i$ ($i=1$ to $N_1$), 
and sheets two and three connected by $N_2$ 
square-root branch-cuts centered about $x=\hat{e}_i$ ($i=1$ to $N_2$), 
which is a Riemann surface of genus $N_1+N_2-2$.   

We rewrite the cubic curve (\ref{four}) as
\be
y^3\,+2A(x)\,y^2\,+\,B(x)\,y+\epsilon(x)=\,0,\label{eight}
\ee
where 
\be
\epsilon(x)\,=-L_1^4\,L_2^2\,P_0^2(x)\,P_3(x)\,\,\,\,, \hs{9mm}
A(x)= - \frac{1}{2}P_1(x)\,\,\,\,,\hs{9mm} 
B(x)= L_1^2 P_0(x)P_2(x).\label{nine}
\ee
As in our previous work 
\cite{NaculichRhedinSchnitzer,EnnesNaculichRhedinSchnitzer}, 
we will solve the prepotential for this 
problem by means of a systematic expansion in powers of $\ep$, with 
the zeroth-order term being a hyperelliptic curve. 
The solutions to (\ref{eight}), correct to ${\cal O}(\ep)$, are 
\be
y_1\,= -\,A\,-\,r\,-{\epsilon\over {2r(A+r)}}\,\,\,\,,\hs{9mm}
y_2\,= - \,A\,+\,r\,+{\epsilon\over {2r(A-r)}}\,\,\,\,,\hs{9mm}
y_3\,=\,-{\epsilon\over B}.\label{ten}
\ee
where $r\equiv\sqrt{A^2-B}$.
Notice that to this order, only sheets labelled by $y_1$ and $y_2$ 
are connected by branch-cuts, while $y_3$ is disconnected. 
(However, the involution 
map (\ref{five}) will enable us to discuss the effects of the connection of 
sheets $y_2$ and $y_3$ by branch-cuts.) 

The SW differential is
\be
\la=x\frac{{\rm d}y}{y},
\ee
which takes a different value on each of the Riemann sheets. 
The perturbative expansion in $\ep$, (\ref{ten}), induces a comparable 
expansion for the SW differential. For example, on sheet one 
\be
\lambda_1\,=\,(\lambda_1)_{I}\,+(\lambda_1)_{II}+...\label{eleven}
\ee
where 
\be
(\lambda_1)_{I}\,=\,{\rm d}x\,x \left({({A'\over A}-{B'\over {2B}})
\over{\sqrt{1-{B\over A^2}}}}\,
+\,{B'\over 2B} \right),\label{twelve}
\ee
is the usual expression for the SW differential for a 
hyperelliptic curve, while the ${\cal O}(\ep)$ correction is 
\be
(\lambda_1)_{II}\,=\,{\rm d}x\,x\,\pr_x\left({\epsilon\over {2Br}}+
{\epsilon\,r\over 
{B^2}}\right).\label {thirteen}
\ee
Equation (\ref{five}) maps the sheets as follows: $y_1\llra y_3$ and 
$y_2\llra y_2$. Using $y_3=L_1^2L_2^2P_0P_3/y_1$, we may 
express the expansion  for $\la_3$ in terms of a comparable one for $\la_1$, 
for which SU$(N_1)\llra $ SU$(N_2)$, 
with the approximation (\ref{ten}) exhibiting the branch-cuts which connect 
sheets 2 and 3. 

Given the SW differential to the required accuracy,
we are able to compute the order parameters and 
dual order parameters to the comparable order in $\ep$. 
Define a set of canonical homology cycles $A_k$ and $B_k$ for Riemann 
sheets $y_1$ and $y_2$, 
and cycles $\hat{A}_k$ and $\hat{B}_k$ for Riemann 
sheets $y_2$ and $y_3$. 
The cycle $A_k$ is chosen to be a simple contour 
enclosing the slit centered about $e_k$ ($k=1$ to $N_1$) on 
sheet 1, while $\hat{A}_k$ ($k=1$ to $N_2$) similarly encloses 
the slit centered about $\hat{e}_k$ on sheet 3. 
Then 
\be
2\pi i a_k=\oint_{A_k}\la_1 \hs{15mm} {\rm and} \hs{15mm}
2\pi i \hat{a}_k=\oint_{\hat{A}_k}\la_3. 
\ee
A calculation essentially identical to that of 
ref. \cite{DHokerKricheverPhong1} or of sec. 4 of 
ref. \cite{NaculichRhedinSchnitzer,EnnesNaculichRhedinSchnitzer} gives
\be
a_k\,&=&\,e_k\,+
\,{1 \over 4} L_1^2 \,{\pr S_k\over \pr x}(e_k) + \cdots
\hs{10mm} (k=1\,\, {\rm to}\,\,N_1), \cr
\hat a_k\,&=&\,\hat{e}_k\,+
\,{1 \over 4} L_2^2 \,{\pr \widehat S_k\over \pr x}(\hat{e}_k)+ \cdots
\hs{10mm} (k=1\,\, {\rm to}\,\,N_2).\label{seventeen}
\ee
where
\be
& &S_k(x)\,=\,{4 x^{K_0}\,\prod_{i=1}^{N_2}\,(x-\hat{e}_i)\over 
{\prod_{i\neq k}^{N_1}\,(x-e_i)^2}}\hs{10mm} (k=1\,\, {\rm to}\,\, N_1)\\
& &\widehat S_k(x)\,=\,{4x^{K_3}\,\prod_{i=1}^{N_1}\,(x-e_i)\over 
{\prod_{i\neq k}^{N_2}\,(x-\hat{e}_i)^2}} \hs{10mm}(k=1\,\, {\rm to}
\,\, N_2).
\label{sixteen}
\ee

We may also compute the dual order parameters, 
\be
2\pi i a_{D,k}=\oint_{B_k}\la_1 \hs{15mm} {\rm and} \hs{15mm}
2\pi i \hat{a}_{D,k}=\oint_{\hat{B}_k}\la_3. 
\ee
where the $B_k$ are curves going 
from $x_1^-$ to $x_k^-$ on the first sheet and from 
$x_k^-$ to $x_1^-$ on the second. 
(The cut centered about $e_k$ goes from $x^-_k$ to $x^+_k$.) 
Analogously, $\hat{B}_k$ are cycles which 
go from sheet 2 to 3. 
The branch cuts $x_k^-$ are computed as in 
refs. \cite{NaculichRhedinSchnitzer} and \cite{EnnesNaculichRhedinSchnitzer}.
A calculation along the lines of sec. 5 
of ref. \cite{NaculichRhedinSchnitzer} gives, 
including the ${\cal O}(\ep)$ 
correction to $\la$, 
\be
2\pi i\,a_{D,k}&=&\,[2N_1\,-\,N_2\,-\,K_0\,+\,2\,{\rm log}\,(-L_1)]\,a_k \nn \\
&-&2\,\sum_{j\neq k}^{N_1}\,(a_k- a_j)\,{\rm log}\,(a_k-a_j)\,
+\sum_{i=1}^{N_2}\,(a_k-\hat a_i)\,{\rm log}\,(a_k-\hat a_i)
+\,K_0\,a_k\,{\rm log}\,a_k \nn \\
&+&L_1^2\left[-\frac{1}{2}\sum_{j=1}^{N_1}\frac{\pr S_j}{\pr x}(a_j)+
\frac{1}{4}\frac{\pr S_k}{\pr x}(a_k)-\frac{1}{2}\sum_{i\neq k}^{N_1}
\frac{S_i(a_i)}{a_k-a_i}\right]\nn \\
&+&L_2^2\left[\frac{1}{4}\sum_{j=1}^{N_2}\frac{\pr \hat S_j}{\pr x}(\hat a_j)
+\frac{1}{4}\sum_{i=1}^{N_2}
\frac{\hat S_i(\hat a_i)}{a_k-\hat a_i}\right]+...  \label{dordint}
\ee
Considerations analogous to those of Appendix D of ref. 
\cite{NaculichRhedinSchnitzer} give us the identities
\be
\sum_{j=1}^{N_1}\frac{\pr S_j}{\pr x}(e_j)=0,\,\,\,\,\,\,\,\,\,\,\,
\sum_{j=1}^{N_2}\frac{\pr \hat S_j}{\pr x}(\hat e_j)=0,  \label{vaninshid}
\ee
implying $ \sum_{i=1}^{N_1}a_i=\sum_{i=1}^{N_1}e_i $
and $ \sum_{i=1}^{N_2} \hat a_i=\sum_{i=1}^{N_2} \hat e_i $
to the order that we are working. 
Combining eqs. (\ref{dordint}), (\ref{vaninshid}), and identities 
analogous to (6.8) of ref. \cite{NaculichRhedinSchnitzer} 
gives
\be
& &2\pi i\,a_{D,k}=\,[2N_1\,-\,N_2\,-\,K_0\,
+\,2\,{\rm log}\,(-L_1)]\,a_k \nn \\
& &-2\,\sum_{j\neq k}^{N_1}\,(a_k- a_j)\,{\rm log}\,(a_k-a_j)\,
+\sum_{i=1}^{N_2}\,(a_k-\hat a_i)\,{\rm log}\,(a_k-\hat a_i) 
+\,K_0\,a_k\,{\rm log}\,a_k \nn \\
& &+\,{L_1^2\over 4}\,{\pr \over \pr a_k}\sum_{i=1}^{N_1}\,S_i(a_i)\,+
{L_2^2\over 4}\,{\pr \over \pr a_k}\sum_{i=1}^{N_2}\,
\widehat S_i(\hat a_i)
\hs{12mm} (k=1\,\,{\rm to}\,\, N_1).\label{eightteen} 
\ee
Using the involution map (\ref{five}) we obtain $2\pi i \hat{a}_{D,k}$ from 
(\ref{eightteen}) with the substitutions 
\be
a_i\llra \hat{a}_i\,\,\,\,\,\,\,\,\,\,\,\,\,
L_1\llra L_2, \,\,\,\,\,\,\,\,\,\,\,\,\,
K_0\llra K_3 \,\,\,\,\,\,\,\,\, 
S_i\llra\widehat{S}_i. 
\ee
The prepotential, which satisfies 
\be
a_{D,k}=\frac{\pr {\cal F}}{\pr a_k} \hs{5mm} 
(k=1\,\,{\rm to}\,\, N_1),  \,\,\,\,\,\,\,\,\,\,\,\,\,
\hat{a}_{D,k}=\frac{\pr {\cal F}}{\pr \hat{a}_k} \hs{5mm} 
(k=1\,\,{\rm to}\,\, N_2),  
\ee 
is
\be
& &{\cal F}_{1-loop}={i\over 8\pi}\left[\sum_{i,j=1}^{N_1}\,
(a_i-a_j)^2\,{\rm log}\,(a_i-a_j)^2+
\sum_{\alpha,\beta=1}^{N_2}\,
(\hat a_{\alpha}-\hat a_{\beta})^2\,{\rm log}\,(\hat a_{\alpha}-
\hat a_{\beta})^2\,\right.\cr
& &-\left.\sum_{i=1}^{N_1}\,\sum_{\alpha=1}^{N_2}\,
(a_i-\hat a_{\alpha})^2\,{\rm log}\,(a_i-\hat a_{\alpha})^2-
{K_0}\sum_{i=1}^{N_1}\,a_i^2\,{\rm log}\,a_i^2\,-
{K_3}\sum_{\alpha=1}^{N_2}\,\hat a_{\alpha}^2\,{\rm log}\,
{\hat a_{\alpha}^2} \right], \label{oneloop}
\ee
and
\be
2\pi i{\cal  F}_{1-inst}\,=\,{1\over 4}L_1^2\,\sum_{i=1}^{N_1}\,S_i\,(a_i)\,+
\,{1\over 4} L_2^2\,\sum_{i=1}^{N_2}\,\widehat S_i\,(\hat a_i),\label{twenty}
\ee
where 
\be
S_k(a_k) &=&\frac{4a_k^{K_0}\prod_{i=1}^{N_2}(a_k-\hat{a}_i)}{
\prod_{i\neq k}^{N_1}(a_k-a_i)^2} \hs{12mm} (k=1\,\,{\rm to}\,\, N_1),
\cr
\widehat{S}_k(\hat{a}_k)
&=&\frac{4\hat{a}_k^{K_3}\prod_{i=1}^{N_1}(
\hat{a}_k-a_i)}{
\prod_{i\neq k}^{N_2}(\hat{a}_k-\hat{a}_i)^2} \hs{12mm} 
(k=1\,\,{\rm to}\,\, N_2). \label{stwo}
\ee
Since $S_k(a_k)$ and $\widehat S_k(\hat a_k)$ 
depend on both $a_i$ and  $\hat{a}_i$, 
eq. (\ref{twenty}) is {\it not} just the 
naive sum of instanton contributions from each subgroup. 

The one-loop prepotential (\ref{oneloop}) 
agrees with the perturbation theory result.
Further, we have one check 
available from the work of D'Hoker and Phong 
(see the last paper in ref. \cite{DHokerKricheverPhong2}), 
who consider various decoupling limits for 
N=2 SU($N$) gauge theory with a massive 
hypermultiplet in the adjoint representation. In their eq. (6.17) ff. 
they give the one-instanton correction for SU($N$)$\times$SU($N$) 
theory with a bifundamental representation, $K_0=K_3=0$, and a 
single quantum-scale. Our result (\ref{twenty}-\ref{stwo}) 
agrees with theirs, up to an overall constant for the quantum scales.

\noindent{\bf 3. SU($N$) with a symmetric tensor flavor and $N_f$ fundamentals}

In this section, 
we derive the one-instanton contribution 
to the prepotential for 
the N=2 SU($N$) gauge theory
with one matter hypermultiplet in 
the rank two symmetric tensor representation
(with mass $m$)
together with $N_f$ matter hypermultiplets in the defining representation
(with masses $m_k$, $k=1, \cdots, N_f$).
Asymptotic freedom restricts $N_f$ to be less than $ N-2$.
(In previous work
\cite{EnnesNaculichRhedinSchnitzer}, 
we obtained the one-instanton prediction for $N_f=0$.)
The Seiberg-Witten curve for this theory, 
constructed by 
Landsteiner, Lopez, and Lowe \cite{LandsteinerLopezLowe}, 
is
\be
y^3\,+f(x)\,y^2\,+\bLa^2\,x^2\,j(x)\,f(-x)\,y+\,
\bLa^6\,x^6\,j^2(x)\,j(-x)\,
=\,0,\label{aone}
\ee
where
\be
f(x)=\prod_{i=1}^N\,(x-e_i) \,,\hs{6mm}
j(x)=\prod_{j=1}^{N_f}\,(x-f_j)\,,\hs{6mm}
{\rm and} \hs{6mm} \bLa^2=\La^{N-2-N_f}.
\label{atwo}
\ee
The $e_i$  parametrize the classical moduli space, 
and the $f_k$ are related to the masses via $ f_k = {1\over 2}m -m_k$.
(We independently derived the curve for this theory
for $m=m_k=0$  using R-symmetry, and checking against M-theory.) 
The curve (\ref{aone})  has the involution 
\be
y\rightarrow {\bLa^4x^4j(x)j(-x)\over y}\,\,\,\,,\,\,\,
\,\,\,\,\,\,
x\rightarrow -x.\label{athree}
\ee

We begin by calculating the prepotential when the 
symmetric hypermultiplet is massless ($m=0$).
First, we define the residue function 
\be
S_k(x)\,=\,{4\,(-1)^N\,x^2\prod_{j=1}^{N_f}(x-f_j)\prod_{i=1}^N\,(x+e_i)\over 
\prod_{i\neq k} (x-e_i)^2}.
\label{afour}
\ee
A calculation along the lines of refs.
\cite{DHokerKricheverPhong1,NaculichRhedinSchnitzer, EnnesNaculichRhedinSchnitzer} 
gives the renormalized order parameters 
\be
a_k\,=\,e_k\,+{\bLa^2\over 4}{\pr\,S_k\over \pr x}(e_k)+...
\,\,\,\,\,\,\,,\,\,\,\,\,\,\,\,\,\,\,\,\,\,\,\,
\,\,\,\,(k=1\,\,{\rm to}\,\,N).\label{afive}
\ee
The dual order parameters can then be computed in terms of $a_k$ as
\be
& &2\pi i\,a_{D,k}
=\,[N-N_f-2+2\, {\rm log}\,\bLa +(N+2)\,{\rm log}\,(-1)]\,a_k \nn \\
& &
-2\,\sum_{j\neq k}\,(a_k- a_j)\,{\rm log}\,(a_k-a_j)\,
+\sum_{i=1}^{N}\,(a_k+a_i)\,{\rm log}\,(a_k+a_i) \nn \\
& &+\,2\,a_k\,{\rm log}\,a_k
+\,\sum_{i=1}^{N_f}\,(a_k- f_i)\,{\rm log}\,(a_k-f_i)\,+
\,{\bLa^2\over 4}\,{\pr \over \pr a_k}\,\sum_{i=1}^{N}\,S_i(a_i).
\label{asix}
\ee 
This enables us to integrate (\ref{adfrel}), and obtain 
the instanton expansion in (\ref{prepot}), accurate to one-instanton. 
We find that ${\cal F}_{1-loop}$ agrees with perturbation theory, 
and
\be
2\pi i\,{\cal F}_{1-inst}\,=\frac{1}{4}\sum_{i=1}^{N}\,S_i(a_i),\label {aseven}
\ee
where
\be
S_k(a_k)=
\frac{4(-1)^Na_k^2\prod_{j=1}^{N_f}(a_k-f_j)
\prod_{i=1}^N(a_k+a_i)}{\prod_{i\neq k}(a_k-a_i)^2}. \label{ssym}
\ee
Equation (\ref{ssym}) reduces to our previous result 
\cite{EnnesNaculichRhedinSchnitzer} 
for $N_f=0$.

For a symmetric hypermultiplet with mass $m$, 
one shifts $a_k \to a_k + {1\over 2}m$ in ${\cal F}(a)$. 
Thus, eq. (\ref{aseven}) remains valid, but with 
\be
S_k(a_k)=
\frac{4(-1)^N(a_k+m/2)^2\prod_{j=1}^{N_f}(a_k+m_j)
\prod_{i=1}^N(a_k+a_i+m)}{\prod_{i\neq k}(a_k-a_i)^2}. \label{massssym}
\ee
Equation (\ref{massssym}) has all the required double-scaling limits 
as $m$ or $m_k\lra\infty$.

\noindent{\bf 4. 
SU($N$) with an antisymmetric tensor flavor and $N_f$ fundamentals}

In this section, 
we derive the one-instanton contribution 
to the prepotential for 
the N=2 SU($N$) gauge theory
with one matter hypermultiplet in the rank two
 antisymmetric tensor representation
(with mass $m$)
together with $N_f$ matter hypermultiplets in the defining representation
(with masses $m_k$, $k=1, \cdots, N_f$).
Asymptotic freedom restricts $N_f$ to be less than $ N+2$.
(The one-instanton prediction for $N_f=0$ 
was presented in ref.~\cite{NaculichRhedinSchnitzer}.)

The Seiberg-Witten curve for this theory was  derived by 
Landsteiner, Lopez, and Lowe \cite{LandsteinerLopezLowe}.
Through a redefinition of $y$, their curve can be written as
\be
y^3+[x^2f(x)+xB+3A]y^2\,+
\,\bLa^2 j(x)[x^2 f(-x)-xB+3A]y\,+
\bLa^6\,j^2(x) j(-x)\,=\,0\,,\label{fsix}
\ee
where $f(x)$ and $j(x)$ are defined in (\ref{atwo}), and\footnote{
Our definitions of $A$ and $B$ differ from  
ref. \cite{LandsteinerLopezLowe} by the signs in front
of the $f_j$.
These changes are necessary for the one-instanton contribution to
the prepotential to make sense (no logarithmic dependence),
and to agree with known results in overlapping cases (see below).}
\be
\bLa^2\,=\,\La^{N+2-N_f}\,\,\,\,\,\,,\,\,\,\,\,\,\,\,\,\,\,\,
A\,=\,\bLa^2\,\prod_{j=1}^{N_f}\,(-f_j)\,\,\,\,\,\,,\,\,\,\,\,\,\,\,\,\,\,\,
B\,=\,\bLa^2\,\sum_{j=1}^{N_f}\,\prod_{l\neq j}\,(-f_l).\label{fseven}
\ee
The curve (\ref{fsix}) has the involution 
\be
y\rightarrow {\bLa^4\,j(x)\,j(-x)\over y}
\,\,\,\,,\,\,\,\,\,\,\,\,\,\,\,\,\,x\rightarrow-x.\label{feight}
\ee
When $N_f=0$, 
eq. (\ref{fseven}) implies $A = \bLa^2$, $B= 0$,
and the curve (\ref{fsix}) reduces to that given in  \cite{LL}.
\

When the matter hypermultiplets are massless ($m=0$, $m_k=0$),
the form of the curve simplifies. 
For $N_f=1$,
eq. (\ref{fseven}) yields $A = 0$, $B= \bLa^2$.
For $N_f \geq 2$,
eq. (\ref{fseven}) yields $A = B= 0$, and the curve simplifies to
\be
y^3+x^2 f(x) y^2\,+
\,\bLa^2 x^{N_f+2} f(-x)y\,+
\bLa^6\,(-1)^{N_f} x^{3 N_f}=\,0.
\label{fus}
\ee
We independently derived the curve (\ref{fus}).

First, we calculate the prepotential for a massless antisymmetric flavor
($m=0$).
We introduce the residue functions
\be
& & S_k(x)=\frac{4(-1)^N \prod_{j=1}^{N_f}\,(x-f_j)\,\prod_{i=1}^N\,(x+e_i)}
{x^2\,\prod_{i\neq k}\,(x-e_i)^2}  \\
& & S_0(x)=\frac{4(-1)^N\prod_{j=1}^{N_f}\,(x-f_j) \prod_{i=1}^N(x+e_i)}
{\prod_{i=1}^N(x-e_i)^2} \\
& & \bLa^2 R_k(x)=\frac{3 A + B x}{x^2\prod_{i\neq k}(x-e_i)} 
\ee
where, as before,
the $e_i$  parametrize the classical moduli space, 
and $ f_k = {1\over 2}m -m_k$.
The renormalized order parameters are calculated to be
\be
a_k\,=\,e_k\,+
\,\bLa^2 \left({1\over 4}\,{\pr S_k\over \pr x}(e_k)\,-\,R_k(e_k)\right)+...
\,\,\,\,\,\,\,,\,\,\,\,\,\,\,\,\,\,\,\,\,\,\,\,
\,\,\,\,(k=1\,\,{\rm to}\,\,N).\label{fnine}
\ee
Following the strategy of refs.
\cite{NaculichRhedinSchnitzer, EnnesNaculichRhedinSchnitzer},
one obtains the dual order parameters 
\be
& &2\pi i\,a_{D,k}=\,[N-N_f+2+2\, {\rm log}\,\bLa
+(N+2)\,{\rm log}\,(-1)]\,a_k \nn \\
& &
-2\,\sum_{j\neq k}\,(a_k- a_j)\,{\rm log}\,(a_k-a_j)\,
+\sum_{i=1}^{N}\,(a_k+a_i)\,{\rm log}\,(a_k+a_i) 
-\,2\,a_k\,{\rm log}\,a_k\nn \\
& &+\,\sum_{i=1}^{N_f}\,(a_k- f_i)\,{\rm log}\,(a_k-f_i)\,+
\,\bLa^2\,{\pr \over \pr a_k}\, 
\left({1\over 4}\,\sum_{i=1}^N\, S_i(a_i)\,- {1\over 2}\,S_0(0) \right).
\label{ften}
\ee 
Integrating this expression, 
we find that the  one-loop prepotential ${\cal F}_{1-loop}$ 
agrees with the perturbation theory, and that 
the one-instanton contribution to the prepotential is given by
\be
2\pi i{\cal F}_{1-inst.}\,=\,{1\over 4}\,\sum_{k=1}^N\, S_k(a_k)\,-
{1\over 2}\,S_0(0),\label{feleven}
\ee
where
\be
S_k(a_k)
 &=& \frac{4(-1)^N \prod_{j=1}^{N_f}\,(a_k-f_j)\,\prod_{i=1}^N\,(a_k+a_i)}
{a_k^2\,\prod_{i\neq k}\,(a_k-a_i)^2},\cr
S_0(0)
&=& \frac{4(-1)^N\prod_{j=1}^{N_f}\,(-f_j) }
{\prod_{i=1}^N\,a_k}\,\,.\label{fthirteen}
\ee
Equations (\ref{feleven}-\ref{fthirteen}) 
reduce to our previous result 
\cite{NaculichRhedinSchnitzer} 
for $N_f=0$.

The result for an antisymmetric hypermultiplet  with mass $m$
is obtained by shifting $a_i\rightarrow a_i\,+{1\over 2}m$
in ${\cal F}(a)$.
Thus, eq. (\ref{feleven}) remains valid, but with
\be
S_k(a_k)
&=&\frac{4(-1)^N 
\prod_{j=1}^{N_f}\,(a_k+m_j)\,\prod_{i=1}^N\,(a_k+a_i+m)}
{(a_k+ {1\over 2} m)^2\,\prod_{i\neq k}\,(a_k-a_i)^2},\cr
S_0(0)
&=&\frac{4(-1)^N\prod_{j=1}^{N_f}\,(m_j- {1\over 2}m) }
{\prod_{i=1}^N\,(a_k+ {1\over 2}m)}\,\,.\label{ffifteen}
\ee

Equation (\ref{feleven}) may be compared with previously 
available results for SU(2) and SU(3) \cite{DHokerKricheverPhong1}. 
The SU(2) theory with one antisymmetric hypermultiplet of mass $m$
and $N_f$ hypermultiplets in the defining representation with masses $m_k$
is equivalent to SU(2) with $N_f$ hypermultiplets in the defining 
representation with masses $m_k$. 
For $N_c=2$, we have checked that one-instanton prepotential (\ref{feleven}) 
is equal to that given in eq. (4.33b) of \cite{DHokerKricheverPhong1}
for $N_f=1$ 
(with the change of scale $\bLa^2 = {1\over 16} \bar \Lambda_{\rm DKP}^2 $),
and differs by a constant (namely, $-2$ and $m-2 \sum_{k=1}^3 m_k$ respectively)
for $N_f=2$ and $3$ (using $a_1+a_2=0$).
Also, SU(3) with one antisymmetric hypermultiplet of mass $m$
and $N_f$ fundamental hypermultiplets with masses $m_k$
is equivalent to SU(3) with $N_f+1$ defining hypermultiplets with
masses $m^\prime_k$. 
For $N_c= 3$, we found that eq. (\ref{feleven})
is equal to  eq. (4.33b) of \cite{DHokerKricheverPhong1}
for $N_f=1$ and $2$ 
(with the change of scale $\bLa^2 = {1\over 16} \bar \Lambda_{\rm DKP}^2 $
and using $a_1+a_2+a_3=0$),
and differs by a constant 
(namely, $-2$ and $m-2\sum_{k=1}^4 m_k$ respectively) for
$N_f=3$ and $4$, 
provided that 
$m^\prime_k = m_k$ for $k=1,\cdots, N_f$
and $m^\prime_{N_f+1} = -m $.

\vfil\break
\noindent{\bf 5. Concluding remarks}

In this paper, we have derived the one-instanton contribution 
to the prepotential for 
N=2 supersymmetric SU($N$) gauge theory with a matter hypermultiplet 
in the symmetric or antisymmetric representation, together with 
$N_f$ hypermultiplets in the defining representation from the cubic 
non-hyperelliptic curves obtained from M-theory. We also studied the 
comparable expansion for SU$(N_1)\times$SU($N_2$) with one hypermultiplet 
in the bifundamental representation, and $K_0$ and $K_3$ massless 
hypermultiplet in the defining representations of SU($N_1$) and SU($N_2$) 
respectively, using the non-hyperelliptic curve derived from M-theory 
by Witten \cite{Witten1}. This latter calculation makes essential use of 
the involution map (\ref{five}) in order to obtain  the contributions of 
the complete set of periods to the prepotential. Thus, the 
techniques of section two of this paper generalizes the methods 
described in refs. \cite{NaculichRhedinSchnitzer} and 
\cite{EnnesNaculichRhedinSchnitzer}. 

What is striking about the one-instanton contribution 
to the prepotential ${\cal F}_{1-inst.}$
obtained from our work 
and that of DKP \cite{DHokerKricheverPhong1,DHokerKricheverPhong2} 
is its remarkable universality of form
when expressed in terms of the renormalized
order parameters $a_k$. 
In all cases one can express it as
\be
2\pi i {\cal F}_{1-inst.}=\frac{1}{4}\sum_{k=1}^NS_k(a_k)
                         -\frac{1}{2} S_0(0),
\label{universal2}
\ee
where the particular group and representation content appear only in 
the form of the residue functions $S_k(a_k)$ and $S_0(0)$,
and the latter depend only on the leading order (in $L$)
coefficients of the hyperelliptic approximation.
($S_0(0)$ vanishes if there is no second order pole at the origin.)
This universality of form is badly hidden
when ${\cal F}_{1-inst.}$ is expressed 
in terms of SU($N$)-invariant moduli. 
It is our opinion that the universality of form of ${\cal F}_{1-inst.}$ 
when expressed in terms of the renormalized order parameters
has not been adequately explained 
(see, however, 
ref. \cite{IsidroEdelsteinMarinoMas} for progress on this issue).

\vs{2mm}

\noindent{\bf{Acknowledgement:}} We would like to thank 
\"Ozg\"ur Sar{\i}o\~{g}lu for valuable discussions. 
HJS wishes to thank the Physics Department of Harvard
University for their hospitality during the spring semester of 
1998.
SGN thanks the Institute for Theoretical Physics for hospitality
extended to him while this work was being completed.

\end{document}